\documentclass[pra,twocolumn,showpacs,preprintnumbers,amsmath,amssymb,superscriptaddress]{revtex4}
\usepackage{graphicx}% Include figure files
\usepackage{dcolumn}% Align table columns on decimal point
\usepackage{bm}% bold math%\documentclass[aps,prl,twocolumn,groupedaddress]{revtex4}
\usepackage[dvips]{epsfig}

\begin{document}

%\preprint{}


\title{Comparison of quantum discord and relative entropy in some bipartite quantum systems }

\author{M.~Mahdian}
\altaffiliation{%
Author to whom correspondence should be addressed; electronic
mail:  mahdian@tabrizu.ac.ir}
\affiliation{%
Faculty of Physics, Theoretical and astrophysics department , University of Tabriz, 51665-163 Tabriz, Iran}

\author{M. B. ~Arjmandi}

\affiliation{%
Faculty of Physics, Theoretical and astrophysics department , University of Tabriz, 51665-163 Tabriz, Iran}
%
%


%\date{\today}

\begin{abstract}

The study of quantum correlations in High-dimensional bipartite systems is crucial for the development of quantum computing.
We propose relative entropy as a distance measure of correlations may be measured by means of the distance from the quantum state to the closest classical-classical state. In particular, We establish relations between relative entropy and  quantum discord quantifiers obtained by means
of orthogonal projection measurements. We show that for symmetrical X-states density matrices the quantum discord is equal to relative entropy. At the end of paper, various examples of X-states such as two-qubit and qubit-qutrit have been demonstrated.

\end{abstract}

%


%\keywords{Biexciton, Binding energy, Built-in electric field, Two photon absorption}


\maketitle

\section{Introduction}

The main goal of quantum information theory is quantifying and
describing quantum processes and rely on quantum correlations
[1,2,3,4]. These correlations are
essential resources in information and computation sciences and
different measures have been put forward for it. Entanglement as the
cornerstone of correlation measures has an effective role in
information processing and has applications in the quantum
computing, cryptography, superdense coding [5,6]
and has been used to quantify quantum teleportation
\cite{Bennett2,Oh}. Apart from entanglement, quantum states can
exhibit other correlations not present in classical systems and it could be a new resource for quantum computation. In order to quantify quantum correlations, the suitable measure is so called the
quantum discord, introduced by Olliver and Zurek \cite{zurek} and
also by Henderson and Vedral \cite{Vedral} independently. Over the past decade, quantum discord has received a lot of attention and many studies performed and articles written about that \cite{ Li, Lang,
Mahdian, Luo, Girolami1, Daki, Girolami2, Saif, Shunlong}. For pure states, quantum discord reduces to entanglement but has a nonzero
value for some mixed separable states and define as the discrepancy between total correlation and classical correlation. However, for mixed quantum states, evaluation of quantum discord is based on minimization procedures over all possible positive operator valued measures (POVM), or von Neumann measurements that can be performed on the subsystems and thus it is somewhat difficult to calculate even numerically.\\
Quite recently, a few analytical results of quantum discord including especially the case of two qubits states such as  rank-2 states \cite{shi} and Bell-diagonal \cite{Luo} have been obtained. In addition, for a rather limited set of two-qubit states, the so-called X states, an analytical formula of quantum discord is proposed by Ali et al \cite{mazhar}. We know that quantum discord measures the amount of information that
cannot be obtained by performing the measurement on one subsystem alone and state after measurement is conditional state. But, in this paper we consider  some density matrices that after performing measurement over all the subsystems are  conditional density matrix that are classical-classical state  (which means that right and left quantum discord are equal) . So, for these kind of quantum states quantum discord is equal to relative entropy and calculation of relative entropy would be too easy.\\
\emph{Quantum discord}. The classical mutual information $I(A:B)$ for  two discrete random variables A and B, is defined as $I(A:B)=H(A)+H(B)-H(A,B)$. Here, $H(p)=-\sum_{i}p_{i} \log p_{i} $ denotes the shannon entropy of the proper distribution \cite{Nielsen}. For a classical
probability distribution, Bayes' rule: $p(a_{i},b_{j})=p(a_{i}|b_{j})p(b_{j})=p(b_{j}|a_{i})p(a_{i})$,
leads to an equivalent definition of the mutual information as
$I(A:B)=H(A)-H(A|B)$.\\
For a given quantum density matrix of a composite system $\rho_{AB}$ , the total amount of correlations, including classical and quantum correlations, is quantified by the quantum mutual information as
\begin{equation}
I(\rho_{AB})=S(\rho_{A})+S(\rho_{B})-S(\rho_{AB}),
\end{equation}
where $S(\rho) = -Tr(\rho log \rho)$ denotes the von Neumann entropy of the relevant state and suppose A and B share a quantum state $\rho_{AB}\in
\cal{H}_{A}\otimes \cal{H}_{B}$.

Assume that we perform a set of local projective measurements (von Neumann measurements)
$\{\Pi^{(j)}_{B}=|j_B\rangle\langle j_B|\}$ on subsystem B .

The measurements will disturb subsystem B and the whole system AB simultaneously.

If the measurement is taken over all
possible complete set of von Neumann projective measurement (one-dimensional orthogonal projectors), described
by ${\{\Pi^{j}_{B}}\}$, corresponding to outcomes j, on subsystem B the resulting
state is given by the shared ensemble $\{\rho_{A|i}, P_{i}\}$, where $ \rho_{A|i} $ is conditional density matrix of bipartite system :
\begin{equation}
\rho_{A|i}=\frac{1}{p_{i}}{(I_A\otimes\Pi^{j}_{B}) \rho_{AB} (I_A\otimes\Pi^{j}_{B})},
\end{equation}
and after taking partial trace over subsystem B, the result is state of subsystem A in form :
\begin{equation}
\rho_{A|i}=\frac{1}{p_{i}}\texttt{Tr}_B\{(I_A\otimes\Pi^{j}_{B}) \rho_{AB} (I_A\otimes\Pi^{j}_{B})\},
\end{equation}
\begin{equation}
P_{i}=\texttt{Tr} (\Pi^{j}_{B}\rho_{AB}\Pi^{j}_{B}),
\end{equation}
with $I_A$ being the identity matrix of subsystem A
and $\texttt{Tr}_{B}$ denotes the partial trace of over subsystem B.

As an  example for the set of measurment, for the state of two qubits
\begin{equation}
 \Pi^{1}=\frac{1}{2} (I+\sum_{j} n_{j} \sigma_{j}),
\end{equation}
\begin{equation}
 \Pi^{2}=\frac{1}{2} (I-\sum_{j} n_{j} \sigma_{j}),
\end{equation}
that $ \sigma_{j} $ are the Pauli matrices and $ \widehat{n} $ is the Bloch sphere eigen vectors :
\begin{equation}
\widehat{n}=(\widehat{n}_{x}, \widehat{n}_{y}, \widehat{n}_{z})=(sin\theta cos\phi, sin\theta sin\phi, cos\theta) .
\end{equation}

A quantum analogue of the conditional entropy can then be defined as
$S_{\{\Pi^{j}_{B}\}}(A|B)\equiv\sum_{i}P_{i}S(\rho_{A|i})$ and an
alternative version of the quantum mutual information can now be defined as $J_{\{\Pi^{j}_{B}\}}(A|B)=
S(\rho_{A} )- S_{\{\Pi^{j}_{B}\}}(A|B)$ where $\rho_{A} = \texttt{Tr}_{B}(\rho)$ and $\rho_{B} = \texttt{Tr}_{A}(\rho)$
 are the reduced density matrix of subsystem A and subsystem B. The
above quantity depends on the selected set of von Neumann measurements or a suitable set of orthogonal projectors ${\{\Pi^{j}_{B}}\}$ . To get all the classical correlations present
in $\rho_{AB}$, we maximize $J_{\{\Pi^{j}_{B}\}}(\rho_{AB})$, over
all ${\{\Pi^{j}_{B}}\}$
\begin{equation}\label{122}
J (\rho_{AB})= Max_{\{\Pi^{j}_{B}\}}\{S(\rho_{A})-S_{\{\Pi^{j}_{B}\}}(A|B)\}.
\end{equation}
Then, quantum discord on subsystem B is defined (right quantum
discord) as:
$$D_{R}(\rho_{AB})=I(\rho_{AB})-J(\rho_{AB})$$
\begin{equation}
=S(\rho_{B})-S(\rho_{AB})+Min_{{\{\Pi^{j}_{B}}\}}S_{\{\Pi_{j}\}}(A|B).
\end{equation}
If the measurement is taken over all possible POVMs
${\{\Pi^{j}_{A}}\}$ on subsystem A, the resulting state is given by
the shared ensemble $\{\rho_{B|i}, P_{i}\}$, where $ \rho_{B|i} $ is conditional density matrix of bipartite system :
\begin{equation}
\rho_{B|i}=\frac{1}{p_{i}}{(\Pi^{j}_{A}\otimes I_B) \rho_{AB} (\Pi^{j}_{A}\otimes I_B)},
\end{equation}
and by taking partial trace over subsystem A , the result is state of subsystem B :
\begin{equation}
\rho_{B|i}=\frac{1}{p_{i}}\texttt{Tr}_A\{(\Pi^{j}_{A}\otimes I_B) \rho_{AB} (\Pi^{j}_{A}\otimes I_B)\},
\end{equation}
\begin{equation}
P_{i}=\texttt{Tr}(\Pi^{j}_{A}\rho_{AB}\Pi^{j}_{A}),
\end{equation}
with $I_B$ is the identity matrix of subsystem B
and $\texttt{Tr}_{A}$ denotes the partial trace of over subsystem A and similar to above relation, classical correlation and quantum discord
on subsystem A (left quantum discord) defined as
\begin{equation}\label{222}
J (\rho_{AB})= Max_{\{\Pi^{j}_{A}\}}\{S(\rho_{B})-S_{\{\Pi^{j}_{A}\}}(A|B)\},
\end{equation}
and
\begin{equation}\label{QD}
D_{L}(\rho_{AB})=S(\rho_{A})-S(\rho_{AB})+Min_{{\{\Pi^{j}_{A}}\}}S_{\{\Pi^{j}_{A}\}}(B|A).
\end{equation}
It has been shown that $D_{R}(\rho)$, $D_{L}(\rho)$ are always
non-negative  and is not symmetric, i.e.
$D_{R}(\rho)\neq D_{L}(\rho)$  in general \cite{zurek}.\\
As mentioned above, for quantifying correlations we need to apply optimization over POVM measures to extract all of classical correlations which is a nontrivial task \cite{Nielsen, Brandt, Sandor}. Therefore, it is difficult to calculate quantum discord in the general case
since the optimization should be taken. In this article, for some bipartite density matrices in SU(N) algebra (i.e. X-states), we will see that density matrix of composite subsystems A and B  after von Neumann measurements is conditional density matrix of bipartite system that is classical-classical state and for these X-states right and left quantum discord are equal. Therefor, the set of measurement will be complete and our states change in a classical state, so we just extract classical correlation from it. Maximization of $J(\rho_{AB})$ captures the maximum classical correlation that can
be extracted from the system, and whatever extra correlation that may remain is the quantum correlation. So, for these bipartite quantum systems,  we can use relative entropy of discord instead of quantum discord.\\
The organization of this paper is as follows. In Sec. II we explain
relative entropy of discord and reveal the relation between quantum
discord and relative entropy of discord. In Sec. III, we give an
explanation about SU(N) algebra and general form of the density
matrix are obtained. In Sec. IV, we perform our investigation on two-qubit states and established relations. In Sec. V, we perform our
inquiry on qubit-qutrit states and got the same results. Finally, we
summarize our results in Sec. VI.

\section{Relative Entropy Of Discord}

The relative entropy is a non-negative and appropriate  measure of distance
between two arbitary states, which
is defined as \cite{Modi3}
\begin{equation}
S(\rho\|\gamma)=\texttt{Tr}(\rho log_2\rho-\rho log_2\gamma).
\end{equation}

By using the consept of relative entropy, we can define the Geometric Discord (GD) as the minimum of distance between closest classical-classical state and the state of bipartite system

\begin{equation}
GD_{rel}(\rho_{AB})=Min_{\chi\in \cal{C}}S(\rho_{AB}\|\chi),
\end{equation}

which $\chi $ belongs to set of classical states ($\cal{C}$) and minimum is taken over all possible states $\chi$. \\
 Modi et al. have showed in [29] that definition of relative entropy represented by equation (15) can be replaced with
\begin{equation}
S(x\|y)=S(y)-S(x).
\end{equation}
So we will have

$$GD_{rel}(\rho_{AB})=S(\rho_{AB}\|\chi)=S(\chi_{\rho_{AB}})-S(\rho_{AB})=$$
\begin{equation}\label{REL}
\texttt{Tr}(\rho_{AB}\log\rho_{AB}-\rho_{AB}\log{\chi_{\rho_{AB}}}).
\end{equation}
Moreover the closest classical-classical state is defined by
\begin{equation}
\chi_{\rho_{AB}}=\sum_{j}(\Pi^j_{A}\otimes\Pi^j_{B})\rho_{AB}(\Pi^j_{A}\otimes\Pi^j_{B}),
\end{equation}
where $ \Pi^j $ is the von Neumann projective measurment which acts on subsystems A and B.
Now, it can be shown that after von Neumann measurment the result will be conditional state as the equations (2) and (10) and for this kind of density matrix e.g. X-state, it is classical-classical state. So for these bipartite systems, the optimization problem over measurements that used for computing the quantum discord in equation (14)  can be turn into minimization of distance between the density matrix of bipartite system and closest classical-classical state. Then we can use the geometric discord instead of quantum discord for these density matrices.

Let the projection measurements $\Pi^{1}_{A,B}$ and $\Pi^{2}_{A,B}$ effect on subsystems A and B. After apply the measurements, the conditional states will be
\begin{equation}
\rho_{A|i}=\sum_{i}\frac{(\mathbb{I}\otimes \Pi^{i}_B)\rho_{AB}(\mathbb{I}\otimes \Pi^{i}_B)}{\texttt{Tr} (\Pi^{i}_B\rho_{AB}\Pi^{i}_{B})},
\end{equation}
\begin{equation}
\rho_{B|i}=\sum_{i}\frac{(\Pi^{i}_A\otimes \mathbb{I})\rho_{AB}(\Pi^{i}_A\otimes \mathbb{I})}{\texttt{Tr} (\Pi^{i}_{A}\rho_{AB}\Pi^{i}_{A})}.
\end{equation}
We have investigated  for these bipartite systems that considered here, after projective measurements on subsystems we get conditional states
 $$\rho_{A|i}=\rho_{A|\Pi^{1}_B}+\rho_{A|\Pi^{2}_B}=\chi_{\rho_{AB}},$$
 and also
 $$\rho_{B|i}=\rho_{B|\Pi^{1}_A}+\rho_{B|\Pi^{2}_A}=\chi_{\rho_{AB}},$$
 and here

 $$\texttt{Tr}(\Pi^{i}_{B}\rho_{AB}\Pi^{i}_{B})=\texttt{Tr}(\Pi^{i}_{A}\rho_{AB}\Pi^{i}_{A})=Tr(\chi_{\rho_{AB}})=1. $$

 So, with compression of equations (\ref{REL}) and (\ref{QD}), after some calculation for these bipartite quantum states(X-state) we get
\begin{equation}
S(\chi_{\rho_{AB}})=S(\rho_{B})+Min_{{\{\Pi^{j}}\}}S_{\{\Pi^{j}\}}(B|A).
\end{equation}
So we have

$$D_{R,L}=S(\rho_{B})+Min_{{\{\Pi^{j}}\}}S_{\{\Pi^{j}\}}(B|A)-S(\rho_{AB})$$
\begin{equation}\label{re-dis}
=S(\chi_{\rho_{AB}})-S(\rho_{AB})=GD_{rel}.
\end{equation}

We apply this method for various examples of X-states such as on two qubits and calculate the quantum discord for these quantum states and show that the result is equal with results of previous papers, especially with Mazhar Ali's work in reference [11].\\
One of the other measure of quantum correlation is the quantum deficit which is defined as difference between the work or information of total system and information of subsystems after effect the LOCC operations to localization of information \cite{Modi3}. It is categorized such zero, one and two-way deficit that are different in type of interaction between subsystems. The zero-way quantum deficit is quantified as minimum of distance between the state of  system and classical-classical state
\begin{equation}
\Delta=Min_{\Pi_{a},\Pi_{b}}(S(\chi_{\rho_{AB}})-S(\rho_{AB})),
\end{equation}
where $ \chi_{\rho_{AB}} $ is classical-classical state that represented in equation (19). So by these explanations, the zero-way quantum deficit is equal to minimum of relative entropy.

%%%%%%%%%%%%%%%%%%%%%%%%%%%%%%%%%%%%%%%%%%%%%%%%%%%%%%%%%%%%%%%%%%%%%%%%%%%%%%
\section{SU(N) Description}

In this section, we show Hermitian operator on a discrete
N-dimensional Hilbert space $\cal{H}$ versus generators of the
SU(N) algebra \cite{Schlienz}. To obtain the generators of the SU(N)
algebra, introduce a set of N projection operators as follows:
\begin{equation}
\widehat{P}_{jk}=|j\rangle\langle k|,
\end{equation}
where $|n\rangle$ are the orthonormalized eigenstates of the linear
Hermitian operator. We can make $N^{2}-1$ operators with
\begin{equation}
\widehat{U}_{jk}=\widehat{P}_{jk}+\widehat{P}_{kj},
\end{equation}
\begin{equation}
\widehat{V}_{jk}=-i(\widehat{P}_{jk}-\widehat{P}_{kj}),
\end{equation}
\begin{equation}
\widehat{W}_{l}=\sqrt{\frac{2}{l(l+1)}}(P_{11}+\cdots+P_{ll}-lP_{l+1,l+1}),
\end{equation}

where \,$1\leq j<k\leq N$\,\,\,,\,\,\,$1\leq l\leq N-1$,

the set of the resulting operators are given by
\begin{equation}
\{\widehat{\lambda}_{j}\}=\{\widehat{U}_{jk}\}\cup\{\widehat{V}_{jk}\}\cup\{\widehat{W}_{l}\},
\end{equation}
$$\{j=1,2,...,N^{2}-1\},$$
the matrices $\{\widehat{\lambda}_{j}\}$ are called generalized Pauli
matrices or SU(N) generators and density matrix for this algebra is
represented by
\begin{equation}
\rho=\frac{1}{N}\mathbb{I}+\frac{1}{2}\sum_{j=1}^{N^{2}-1}\lambda_{j}\widehat{\lambda}_{j}.
\end{equation}
They also satisfy the following relations:

$$Tr(\widehat{\lambda}_{i}\widehat{\lambda}_{j})=2\delta_{ij},$$

$$S_{j}=Tr\{\widehat{\lambda}_{j}\rho\},$$

$$Tr\{\widehat{\lambda}_{j}\}=0.$$

For a bipartite system with states $\rho_{AB}\in\cal{H_A}\otimes \cal{H_B},$ $dim H_A=d_{A}$ and $dim H_B= d_{B},$  density matrix is shown in
Fano form \cite{Fano} as
\begin{equation}
\rho_{AB}=\frac{1}{d_{A}d_{B}}(\mathbb{I}_{A}\otimes\mathbb{I}_{B}+
\sum^{N^{2}-1}_{i=1}\alpha_{i}\widehat{\lambda}_{i}^{A}\otimes\mathbb{I}_{B}+
\sum_{j=1}^{N^{2}-1}\beta_{j}\mathbb{I}_{A}\otimes\widehat{\lambda}_{j}^{B}
\end{equation}
\begin{equation}
+\sum_{i=1}^{N^{2}-1}\sum_{j=1}^{N^{2}-1}\gamma_{ij}\widehat{\lambda}_{i}^{A}\otimes\widehat{\lambda}_{j}^{B}).
\end{equation}
Closest classical-classical state with projection operators  $P_{k}=|k\rangle\langle k|$ is given with

$$\chi_{\rho_{(AB)}}=\sum_{k}(P_{k}^{A}\otimes P_{k}^{B})\rho_{AB}
(P_{k}^{A}\otimes
P_{k}^{B})=\frac{1}{d_{A}d_{B}}(\mathbb{I}_{A}\otimes\mathbb{I}_{B}$$
$$+\sum_{k=1}^{N}\sum_{i=1}^{N^{2}-1}\alpha_{i}(P_{k}^{A}\lambda_{i}^{A}P_{k}^{A})\otimes\mathbb{I}_{B}+
\sum_{k=1}^{N}\sum_{j=1}^{N^{2}-1}\beta_{j}\mathbb{I}_{A}\otimes(P_{k}^{B}\lambda_{j}^{B}P_{k}^{B})$$
\begin{equation}
+\sum_{k=1}^{N}\sum_{i=1}^{N^{2}-1}\sum_{j=1}^{N^{2}-1}\gamma_{ij}
(P_{k}^{A}\lambda_{i}^{A}P_{k}^{A})\otimes(P_{k}^{B}\lambda_{j}^{B}P_{k}^{B}),
\end{equation}
where after calculation takes the form:
\begin{equation}
\chi_{\rho_{AB}}=\sum_{A,B}(|k_{A}k_{B}\rangle\langle
k_{A}k_{B}|)\rho_{AB}(|k_{A}k_{B}\rangle\langle k_{A}k_{B}|),
\end{equation}
by applying the projection operators  we obtain
\begin{equation}
P_{k}\{\widehat{V}_{jk}\}P_{k}=0,
\end{equation}
\begin{equation}
P_{k}\{\widehat{U}_{jk}\}P_{k}=0,
\end{equation}
\begin{equation}
P_{k}\{\widehat{W}_{l}\}P_{k}\neq0.
\end{equation}
Density matrix of projective measurements on the subsystem A for
density matrix Eq. (21) is

$$\rho_{B|k_1}=\sum_{k}(P_{k}^{A}\otimes
\mathbb{I})\rho_{AB}(P_{k}^{A}\otimes\mathbb{I})=\frac{1}{d_{A}d_{B}}(\mathbb{I}_{A}\otimes\mathbb{I}_{B}$$
$$+\sum_{k=1}^{N}\sum_{i=1}^{N^{2}-1}\alpha_{i}(P_{k}^{A}\lambda_{i}^{A}P_{k}^{A})\otimes\mathbb{I}_{B}
+\sum_{k=1}^{N}\sum_{j=1}^{N^{2}-1}\beta_{j}\mathbb{I}_{A}\otimes
\lambda_{j}^{B}$$
\begin{equation}
+\sum_{k=1}^{N}\sum_{i=1}^{N^{2}-1}\sum_{j=1}^{N^{2}-1}\gamma_{ij}(P_{k}^{A}\lambda_{i}^{A}P_{k}^{A})\otimes
\lambda_{j}^{B}),
\end{equation}
with considering Eqs. (25, 26, 27) we will gain
$$\rho_{B|k_1}=\frac{1}{d_{A}d_{B}}(\mathbb{I}_{A}\otimes\mathbb{I}_{B}$$
$$+\sum_{k=1}^{N}\sum_{i=1}^{N^{2}-1}\alpha_{i}(P_{k}^{A}\{\widehat{W}_{l}\}P_{k}^{A})\otimes\mathbb{I}_{B}
+\sum_{k=1}^{N}\sum_{j=1}^{N^{2}-1}\beta_{j}\mathbb{I}_{A}\otimes
\lambda_{j}^{B}$$
\begin{equation}
+\sum_{k=1}^{N}\sum_{i=1}^{N^{2}-1}\sum_{j=1}^{N^{2}-1}\gamma_{ij}(P_{k}^{A}\{\widehat{W}_{l}\}P_{k}^{A})\otimes
\lambda_{j}^{B}).
\end{equation}

Density matrix of projective measurements on the subsystem B for
density matrix Eq. (22) is

$$\rho_{A|k_2}=\sum_{k}(\mathbb{I}\otimes P_{k}^{B})\rho_{AB}(\mathbb{I}\otimes
P_{k}^{B})=\frac{1}{d_{A}d_{B}}(\mathbb{I}_{A}\otimes\mathbb{I}_{B}$$
$$+\sum_{k=1}^{N}\sum_{i=1}^{N^{2}-1}\alpha_{i}\lambda_{i}^{A}\otimes\mathbb{I}_{B}+
\sum_{k=1}^{N}\sum_{j=1}^{N^{2}-1}\beta_{j}\mathbb{I}_{A}\otimes
(P_{k}^{B}\lambda_{j}^{B}P_{k}^{B})$$
\begin{equation}
+\sum_{k=1}^{N}\sum_{i=1}^{N^{2}-1}\sum_{j=1}^{N^{2}-1}
\gamma_{ij}\lambda_{i}^{A}\otimes(P_{k}^{B}\lambda_{j}^{B}P_{k}^{B})),
\end{equation}
with considering Eqs. (25, 26, 27) we will get
$$\rho_{A|k_2}=\frac{1}{d_{A}d_{B}}(\mathbb{I}_{A}\otimes\mathbb{I}_{B}$$
$$+\sum_{k=1}^{N}\sum_{i=1}^{N^{2}-1}\alpha_{i}\lambda_{i}^{A}\otimes\mathbb{I}_{B}+
\sum_{k=1}^{N}\sum_{j=1}^{N^{2}-1}\beta_{j}\mathbb{I}_{A}\otimes
(P_{k}^{B}\{\widehat{W}_{l}\}P_{k}^{B})$$
\begin{equation}
+\sum_{k=1}^{N}\sum_{i=1}^{N^{2}-1}\sum_{j=1}^{N^{2}-1}\gamma_{ij}\lambda_{i}^{A}
\otimes(P_{k}^{B}\{\widehat{W}_{l}\}P_{k}^{B})).
\end{equation}

The computation of quantum discord is dependent to optimization of measurment. In the next sections, two examples of X-state density matrix as Two-Qubit and Qubit-Qutrit have been presented. We show that optimization of measurment can be replaced by minimum of distance between the state of bipartite system and its classical-classical state. Also we are following to extend this method to higher bipartite systems.

%%%%%%%%%%%%%%%%%%%%%%%%%%%%%%%%%%%%%%%%%%%%%%%%%%%%%%%%%%%%%%%%%%%%%%%%%%%%%%%%%%
\section{Two-Qubit density matrices}

As the first example, we investigate two qubits state which we frequently encounter in condensed matter systems, quantum dynamic, etc. and apply our achievements. The general form of two qubits density matrix is given by
\begin{equation}\label{density}
\rho_{AB}=\frac{1}{4}(\mathbb{I}_{2}\otimes
\mathbb{I}_{2}+\sum_{i=1}^{3}\alpha_{i}\sigma_{i}\otimes
\mathbb{I}_{2}+\sum_{i=1}^{3}\beta_{i}\mathbb{I}_{2}\otimes
\sigma_{i}+\sum_{i,j=1}^{3}\gamma_{ij}\sigma_{i}\otimes\sigma_{j}),
\end{equation}

where $\alpha_{i}, \beta_{i}, \gamma_{ij}\in\mathbb{R},$\ and
$\sigma_{i}$ $(i=1,2,3)$ are three Pauli matrices and $\mathbb{I}$
is identity matrix. For this density matrix, closest
classical-classical state according Eq.(25) calculate as follow

$$\chi_{\rho_{AB}}=\frac{1}{4}(\mathbb{I}_{2}\otimes\mathbb{I}_{2}
+\sum_{k_{A}=1}^{2}\sum_{i=1}^{3}\alpha_{i}(|k_{A}\rangle\langle
k_{A}|\sigma_{i}^{A}|k_{A}\rangle\langle
k_{A}|)\otimes\mathbb{I}_{2}+$$
$$\sum_{k_{B}=1}^{2}\sum_{j=1}^{3}\beta_{j}\mathbb{I}_{2}\otimes
(|k_{B}\rangle\langle k_{B}|\sigma_{j}^{B}|k_{B}\rangle\langle
k_{B}|)+$$
\begin{equation}
\sum_{k_{A}=1}^{2}\sum_{k_{B}=1}^{2}\sum_{i,j=1}^{3}\gamma_{ij}(|k_{A}\rangle\langle
k_{A}|\sigma_{i}^{A}|k_{A}\rangle\langle
k_{A}|)\otimes(|k_{B}\rangle\langle
k_{B}|\sigma_{j}^{B}|k_{B}\rangle\langle k_{B}|).
\end{equation}
With refer to equations (2) and (10) and apply the measurments on $ \rho_{AB} $, the conditional states obtain as
$$
\rho_{B|k_1}=\frac{1}{4}(\mathbb{I}_{2}\otimes\mathbb{I}_{2}+
\sum_{k_{A}=1}^{2}\sum_{i=1}^{3}\alpha_{i} (|k_{A}\rangle\langle
k_{A}|\sigma_{i}^{A}|k_{A}\rangle\langle
k_{A}|)\otimes\mathbb{I}_{2}$$
\begin{equation}+\sum_{j=1}^{3}\beta_{j}\mathbb{I}_{2}\otimes
\sigma_{j}^{B}+\sum_{k_{A}=1}^{2}\sum_{i=1}^{3}\sum_{j=1}^{3}\gamma_{ij}(|k_{A}\rangle\langle
k_{A}|\sigma_{i}^{A}|k_{A}\rangle\langle k_{A}|)\otimes
\sigma_{j}^{B}),
\end{equation}
and

$$\rho_{A|k_2}=\frac{1}{4}(\mathbb{I}_{A}\otimes\mathbb{I}_{B}
+\sum_{i=1}^{3}\alpha_{i}\sigma_{i}^{A}\otimes\mathbb{I}_{B}$$
$$+\sum_{k_{B}=1}^{2}\sum_{j=1}^{3}\beta_{j}\mathbb{I}_{A}\otimes
(|k_{B}\rangle\langle k_{B}|\sigma_{j}^{B}|k_{B}\rangle\langle
k_{B}|)$$
\begin{equation}
+\sum_{k_{B}=1}^{2}\sum_{i=1}^{3}\sum_{j=1}^{3}\gamma_{ij}\sigma_{i}^{A}\otimes(|k_{B}\rangle\langle
k_{B}|\sigma_{j}^{B}|k_{B}\rangle\langle k_{B}|)).
\end{equation}

We choose the measurment in eigenbasis of $ \sigma_z $ i.e
$$|k_{A}\rangle\langle k_{A}|= |0\rangle\langle 0|,$$
and
$$|k_{B}\rangle\langle k_{B}|= |1\rangle\langle 1|.$$
In this paper, we consider X-state density matrix which because of the visual appearance of these density matrices look like to letter X, then are called by this name. By effect the mentioned measurments in equation (40) it becomes X-state by following conditions
 $$\alpha_{1}=\alpha_{2}=\beta_{1}=\beta_{2}=0,$$
$$\gamma_{31} =\gamma_{13}=\gamma_{32}=\gamma_{23}= 0$$
So density matrix for two qubits in form X-state
obtains as
\begin{equation}
\rho_{AB}=\left(
\begin{array}{cccc}
\rho_{11}&0&0&\rho_{14}\\
0&\rho_{22}&\rho_{23}&0\\
0&\rho_{32}&\rho_{33}&0\\
\rho_{41}&0&0&\rho_{44}\\
\end{array}
\right),
\end{equation}
where
$$\rho_{11}=1+\gamma_{33}+\alpha_{3}+\beta_{3},$$
$$\rho_{22}=1-\gamma_{33}+\alpha_{3}-\beta_{3},$$
$$\rho_{33}=1-\gamma_{33}-\alpha_{3}+\beta_{3},$$
$$\rho_{44}=1+\gamma_{33}-\alpha_{3}-\beta_{3},$$
$$\rho_{14}=\rho^\ast_{41}=\gamma_{11}-i\gamma_{12}-i\gamma_{21}-\gamma_{22},$$
$$\rho_{23}=\rho^\ast_{32}=\gamma_{11}+i\gamma_{12}-i\gamma_{21}+\gamma_{22},$$
and also $ \sum_{i}{\rho_{ii}}=1 $. By apply the measurment in equation (32), closest classical-classical state will be
\begin{equation}
\chi_{\rho_{AB}}=\left(
\begin{array}{cccc}
\rho_{11}&0&0&0\\
0&\rho_{22}&0&0\\
0&0&\rho_{33}&0\\
0&0&0&\rho_{44}\\
\end{array}
\right).
\end{equation}
Moreover the conditional states represented by equations (42) and (43) are
\begin{equation}
\rho_{A|i}=\rho_{A|1}+\rho_{A|2}=\chi_{\rho_{AB}},
\end{equation}
and also
\begin{equation}
\rho_{B|i}=\rho_{B|1}+\rho_{B|2}=\chi_{\rho_{AB}}.
\end{equation}
The von Neumann entropy of $ \chi_{\rho_{AB}} $ is

\begin{equation}
S(\chi_{\rho_{AB}})=-\sum \rho_{ii}\log_{2}\rho_{ii},
\end{equation}
In the other hand $ S_{\Pi^j}(B|A) $ will be [11]
\begin{equation}
S_{\Pi^{j}}(B|A)=-\frac{1+\delta_{z}}{2} log \frac{1+\delta_{z}}{2}
- \frac{1-\delta_{z}}{2} log \frac{1-\delta_{z}}{2},
\end{equation}
where $ \delta_{z} = |(\rho_{11}+\rho_{44})-(\rho_{22}+\rho_{33})| $ and also the state of subsystems A and B respectively is
\begin{equation}
\rho_{A}=\left(
\begin{array}{cccc}
\rho_{11}+\rho_{22}&0\\
0&\rho_{33}+\rho_{44}\\
\end{array}
\right),
\end{equation}
\begin{equation}
\rho_{B}=\left(
\begin{array}{cccc}
\rho_{11}+\rho_{33}&0\\
0&\rho_{22}+\rho_{44}\\
\end{array}
\right),
\end{equation}
then von Neumann entropy of $ \rho_{B} $ obtains as
\begin{equation}
S(\rho_{B})=-((\rho_{11}+\rho_{33}) log (\rho_{11}+\rho_{33}) + (\rho_{22}+\rho_{44}) log (\rho_{22}+\rho_{44})).
\end{equation}
By simplify these equations we get
\begin{equation}
S(\rho_B)+S_{\{\Pi^{j}_{B}\}}(B|A)=-\sum \rho_{ii}\log_{2}\rho_{ii}=S(\chi_{\rho_{AB}}).
\end{equation}
So

$$D(\rho_{AB})=S(\rho_B)+S_{\{\Pi^{j}_{B}\}}(B|A)-S(\rho_{AB})$$
\begin{equation}=S(\chi_{\rho_{AB}})-S(\rho_{AB})=GD(\rho_{AB}).
\end{equation}
It can be shown that if $ \delta_{x} $ be the optimal value, so we  choose the von Neumann measurment in the eigenbasis of $ \sigma_{x} $  i.e.  $ |k_{A}\rangle=\frac{|0\rangle+|1\rangle}{\sqrt{2}} $ , $|k_{B}\rangle=\frac{|0\rangle-|1\rangle}{\sqrt{2}} $ and as well as for $ \delta_{y} $  i.e. $ |k_{A}\rangle=\frac{|0\rangle+i|1\rangle}{\sqrt{2}} $ , $|k_{B}\rangle=\frac{|0\rangle-i|1\rangle}{\sqrt{2}} $
, and the results is same to Mazhar Ali et al. [11].

%%%%%%%%%%%%%%%%%%%%%%%%%%%%%%%%%%%%%%%%%%%%%%%%%%%%%%%%%%%%%%%%%%%%%%%%%%%%%%%%%%%%%%%%%%%%%%%%%%%%%%%%%%%%%
\section{Qubit-Qutrit states}

As the second example, we have generalized our relations for Qubit-Qutrit,
including  $\rho_{AB}\in\cal{H_A}\otimes \cal{H_B},$ $dim H_A=2$ and $dim H_B=3$. So,
density matrix versus SU(N) algebra can be represented as
\begin{equation}
\rho=\frac{1}{6}(\mathbb{I}_{2}\otimes\mathbb{I}_{3}+
\sum_{i=1}^{3}\alpha_{i}\sigma_{i}\otimes\mathbb{I}_{3}+\sum_{i}^{8}\sqrt{3}
\beta_{i}\mathbb{I}_{2}\otimes\lambda_{i}
+\sum_{i}^{3}\sum_{j}^{8}\gamma_{ij}\sigma_{i}\otimes\lambda_{j}),
\end{equation}

where $\alpha_{i}, \beta_{i}, \gamma_{ij} \in \mathbb{R},\,
\sigma_{i} (i=1,2,3)$, are three Pauli matrices and $\lambda_{i}
(i=1,\cdots,8)$, are Gell mann matrices and $\mathbb{I}$ is identity
matrix. We applied the conditions until matrix comes in the form
X-state. These conditions are
as $\{\alpha_{3};\beta_{3};\gamma_{33};\gamma_{38};\gamma_{24};\gamma_{14};\gamma_{25};\gamma_{15}\}\neq0,$ and other coefficients are equal to zero. So, density matrix can be written as
$$
\rho_{AB}=\frac{1}{6}(\mathbb{I}_{2}\otimes\mathbb{I}_{3}+
\alpha_{3}\sigma_{3}\otimes\mathbb{I}_{3}+\sqrt{3}
\beta_{3}\mathbb{I}_{2}\otimes\lambda_{3}+\gamma_{33}\sigma_{3}\otimes\lambda_{3}
$$
\begin{equation}\label{density2}
+\gamma_{38}\sigma_{3}\otimes\lambda_{8}+
\gamma_{24}\sigma_{2}\otimes\lambda_{4}+\gamma_{14}\sigma_{1}\otimes\lambda_{4}+\gamma_{25}\sigma_{2}\otimes\lambda_{5}+\gamma_{15}\sigma_{1}\otimes\lambda_{5}).
\end{equation}
By using Eq. (\ref{REL}) relative entropy of discord for this density matrix
Eq. (\ref{density2}) is equal to
\begin{equation}\label{relt1}
D_{rel}(\rho)=\sum_{i=1}^{6}-(\Phi_{i}\log_{2}\Phi_{i}+\Psi_{i}\log_{2}\Psi_{i}),
\end{equation}
where
$$\Phi_{1,2}=\frac{1}{6}(1-2\beta_{8}\pm \alpha_{3}\mp\frac{2\gamma_{38}}{\sqrt{3}}),$$
$$\Phi_{3,4}=\frac{1}{6}(1+\sqrt{3}\beta_{3}+\beta_{8}\mp \alpha_{3}\mp \gamma_{33}\mp\frac{\gamma_{38}}{\sqrt{3}}),$$
$$\Phi_{5,6}=\Psi_{5,6}=\frac{1}{6}
(1-\sqrt{3}\beta_{3}+\beta_{8}-\alpha_{3}\pm
\gamma_{33}\mp\frac{\gamma_{38}}{\sqrt{3}}),$$
$$\Psi_{1,2}=\frac{1}{36}(6+
3\sqrt{3}\beta_{3}-3\beta_{8}+3\gamma_{33}+3\sqrt{3}\gamma_{38}\pm$$
$$\sqrt{3}[9\beta_{3}^{2}+27\beta_{8}^{2}+36\beta_{8}\alpha_{3}+12\alpha_{3}^{2}+
12((\gamma_{15}+\gamma_{24})^{2}+(\gamma_{14}-\gamma_{25})^{2})$$
$$+3\gamma_{33}^{2}+6\beta_{3}
(3\sqrt{3}\beta_{8}+2\sqrt{3}\alpha_{3}+\sqrt{3}\gamma_{33}-\gamma_{38})$$
$$-6\sqrt{3}\beta_{8}\gamma_{38}-4\sqrt{3}\alpha_{3}\gamma_{38}+\gamma_{38}^{2}+2\gamma_{33}
(9\beta_{8}+6\alpha_{3}-\sqrt{3}\gamma_{38})]^{\frac{1}{2}}),$$
$$\Psi_{3,4}=\frac{1}{36}(6 +
3\sqrt{3}\beta_{3}-3\beta_{8}-3\gamma_{33}-3\sqrt{3}\gamma_{38}\pm$$
$$\sqrt{3}[9\beta_{3}^{2}+27\beta_{8}^{2}-36\beta_{8}\alpha_{3}+12\alpha_{3}^{2}+
12((\gamma_{15}-\gamma_{24})^{2}+(\gamma_{14}+\gamma_{25})^{2})$$
$$+3\gamma_{33}^{2}+6\beta_{3}
(3\sqrt{3}\beta_{8}-2\sqrt{3}\alpha_{3}-\sqrt{3}\gamma_{33}+\gamma_{38})$$
$$+6\sqrt{3}\beta_{8}\gamma_{38}-4\sqrt{3}\alpha_{3}\gamma_{38}+\gamma_{38}^{2}+2\gamma_{33}
(-9\beta_{8}+6\alpha_{3}-\sqrt{3}\gamma_{38})]^{\frac{1}{2}}).$$

It can be seen that the result Eq.(\ref{relt1}) is equal to the result is
obtained for quantum discord qubit-qutrit density matrix. Here we consider the set of measurment in eigenbasis of $ S_{z} $ . To
better illustrate the results for $2\times3$ matrices we consider
the following example \cite{Karpat,kapil,Mazhar}\\

$\rho=\frac{p}{2}(|00\rangle\langle00|+|01\rangle\langle01|+|00\rangle\langle12|+
|11\rangle\langle11|+|12\rangle\langle12|+$
\begin{equation}
|12\rangle\langle00|)+\frac{1 -
2p}{2}(|02\rangle\langle02|+|02\rangle\langle10| +
|10\rangle\langle02|+|10\rangle\langle10|),
\end{equation}
where classical correlation $\chi_{\rho}$ are obtained as follows.
\begin{equation}
\chi_{\rho}=\frac{1}{2} \left(
\begin{array}{cccccc}
p&0&0&0&0&0\\
0&p&0&0&0&0\\
0&0&1-2p&0&0&0\\
0&0&0&1-2p&0&0\\
0&0&0&0&p&0\\
0&0&0&0&0&p\\
\end{array}
\right),
\end{equation}

and we have
\begin{equation}
S(\chi_{\rho_{AB}})=1-2p\log_{2}p-(1-2p)\log_{2}(1-2p).
\end{equation}

%%%%%%%%%%%%%%%%%%%%%%%%%%%%%%%%%%%%%%%%%%%%%%%%%%%%%%%%%%%%%%%%%%%%%%%%%%%%%%%%%%%%%%%%%
\section{Conclusions}

In this paper, we have investigated an analytical method of quantum
discord for some bipartite quantum systems. We represent with
orthogonal projective measurement on the subsystems, the resulting matrix will be classical-classical state and set of measurements will be complete. Thus, for these states we obtain after measurement, the optimization over orthogonal
projective measurements can be turn into minimization of distance between
the state of bipartite system and its closest
classical-classical state. This
means that the relative entropy of discord can be replaced with
quantum discord and we have justified our claim with examples that
have mentioned above. We are going to extend this method for case of high bipartite systems in future.

\section{Acknowledgments}
 This work is published as a part of research project supported by the university of
Tabriz research affairs office.

%%%%%%%%%%%%%%%%%%%%%%%%%%%%%%%%%%%%%%%%%%%%%%%%%%%%%%%%%%%%%%%%%%%%%%%%%%%%%%%%%%%%%%%%%%%%%%%%%%%%%%%%%%%

\end{document}